# Negative longitudinal magnetoresistance in GaAs quantum wells

Jing Xu[1,2], Meng K. Ma[3], Maksim Sultanov[2], Zhi-Li Xiao[1,2,*], Yong-Lei Wang[1,4,*], Dafei Jin[5], Yang-Yang Lyu[1,4], Wei Zhang[1,6,*], Loren N. Pfeiffer[3], Ken W. West[3], Kirk W. Baldwin[3], Mansour Shayegan[3], and Wai-Kwong Kwok[1]

[1]*Materials Science Division, Argonne National Laboratory, Argonne, Illinois 60439, USA*

[2]*Department of Physics, Northern Illinois University, DeKalb, Illinois 60115, USA*

[3]*Department of Electrical Engineering, Princeton University, Princeton, New Jersey 08544, USA*

[4]*Research Institute of Superconductor Electronics, School of Electronic Science and Engineering, Nanjing University, Nanjing 210093, China*

[5]*Center for Nanoscale Materials, Argonne National Laboratory, Argonne, Illinois 60439, USA*

[6]*Department of Physics, Oakland University, Rochester, MI 48309, USA*

*Correspondence to: xiao@anl.gov; ylwang@anl.gov; zwei@anl.gov

**Negative longitudinal magnetoresistances (NLMRs) have been recently observed in a variety of topological materials and often considered to be associated with Weyl fermions that have a defined chirality. Here we report NLMRs in non-Weyl GaAs quantum wells. In the absence of a magnetic field the quantum wells show a transition from semiconducting-like to metallic behaviour with decreasing temperature. We observed pronounced NLMRs up to 9 Tesla at temperatures above the transition and weak NLMRs in low magnetic fields at temperatures close to the transition and below 5 K. The observed NLMRs show various types of magnetic field behaviour resembling those reported in topological materials. We attribute them to microscopic disorder and use a phenomenological three-resistor model to account for their various features. Our results showcase a new contribution of microscopic disorder in the occurrence of novel phenomena. They may stimulate further work on tuning electronic properties via disorder/defect nano-engineering.**

Magnetic field induced resistance change is conventionally termed as magnetoresistance (MR), which is usually related to magnetism and plays crucial roles in applications such as sensors and storage devices [1]. In a single-band nonmagnetic material the semiclassical Boltzmann equation approach gives rise to a magnetic field independent resistivity $\rho_0 \equiv m/e^2 n \tau$, where $e$, $m$, $n$ and $\tau$ are the charge, effective mass, density and relaxation time of the charge carriers [2]. Deviations from a constant resistivity, however, are often reported, e.g., in two-dimensional electron gas, where both positive [3] and negative [4] MRs have been observed when the applied magnetic field $B$ is perpendicular to the current $I$ ($B \perp I$).

With the recent discovery of topological materials, the MR phenomenon has been attracting extensive attention [5-23]. Two of the most remarkable findings are the extremely large MRs for $B \perp I$ [5-8] and the negative longitudinal MRs (NLMRs) for $B // I$ [9-23]. The former can be attributed to the co-existence of high-mobility electrons and holes, resulting in diminishing Hall effect [5,24]. The origin of the NLMR, however, is currently under debate. In disordered systems such as films of topological insulator $Bi_2Se_3$ [21] and Dirac semimetal $Cd_3As_2$ [22], the NLMR is attributed to distorted current paths due to conductivity fluctuations induced by macroscopic disorder as revealed in computer simulations for polycrystalline $Ag_{2\pm x}Se$ samples [25,26]. In contrast, the NLMR in single crystals of these materials [9-17] is often considered as a manifestation of Weyl fermions due to the chirality imbalance in the presence of parallel magnetic and electric fields [27,28]. Recently, theories with topological [29,30] and trivial [30-32] states have been developed to understand the observed NLMR without invoking chiral anomaly. However, some puzzling concerns have been raised on NLMR experiments [33]. For example, numerous non-Weyl materials exhibit NLMRs [20,34-36]. Even within the class of Weyl semimetals, NLMRs are not consistently observed in the same material [37,38]. Furthermore, the reported NLMRs show unpredicted features such as non-monotonic magnetic field behaviour [9,10,16]. More disturbingly, it has been demonstrated that NLMRs can be artificially induced by non-uniform current injection [33,39].

Here we report NLMRs in a non-Weyl material and reveal that microscopic disorder can be their origin. We conducted magnetotransport measurements on GaAs quantum wells, in which microscopic disorder induces novel phenomena such as quantum Hall plateaus [40] and linear magnetoresistance [3]. We observed NLMRs with both monotonic and non-monotonic magnetic field dependence similar to those reported in topological materials. Furthermore, the NLMRs in our quantum wells exhibit an intriguing temperature behavior: they occur at temperatures below 5 K, disappear at intermediate temperatures (> 5 K), and re-appear at high temperatures ($T > 130$ K and up to 300 K). The NLMR is most pronounced at $160 - 180$ K. After excluding experimental artifacts and accounting for the various features in their magnetic field dependence using a simple three-resistor model, we attribute the observed NLMRs to microscopic disorder including impurity and interface roughness.

**Results**

We measured four samples in standard Hall-bar geometry, defined through photolithographic patterning. Three of them (see micrographs in Fig.1a and Fig.S1) are Hall bars on the same wafer and connected to each other, sharing the same applied current and denoted as Samples W1a, W1b and W1c in Fig.S1. The fourth sample (sample W2) is the same Hall bar structure fabricated on a separate wafer. All the samples behave qualitatively the same (see Table S1 for a summary of the characteristic parameters including the low-temperature electron densities for all four samples). Here we focus on results of Sample W1b, with additional data from other samples presented in the supplement.

We obtained data on the magnetic field dependence of the sample resistance $R(B)$ at fixed temperatures and angles between the magnetic field and the current (see Fig.1b and Fig.S9 for the definition of angle $\theta$). At temperatures below 5 K, we observed negative MRs for both $B//I$ ($\theta = 0°$) (Fig.1e) and $B\perp I$ ($\theta = 90°$) (see Fig.S2). At $T > 5$ K MRs become entirely positive for both magnetic-field orientations (see Fig.S2). With further increase in temperature, the MRs for $B//I$ begin to display non-monotonic behaviour, and NLMR behaviour re-emerges at the medium magnetic fields (see Fig.1e for MRs at 133 K). At $T > 145$ K and up to room temperature, the MRs become purely negative, although they are non-monotonic at temperatures between 145 K and 210 K. From the $R(B)$ curves obtained at various temperatures, we construct the temperature dependence of $MR = [R(B) - R_0]/R_0$, where $R_0$ is the sample resistance in the absence of an external magnetic field. The Cartesian plots of the results for Sample W1b and the other three samples are presented in Fig.1d and Fig.S3, respectively. Colour maps that can show both the $MR$'s temperature and magnetic field dependences are given in Fig.S4 for all four samples. These plots and maps clearly show that all samples exhibit similar NLMRs, with some quantitative variation from sample to sample, e.g., the maximum NLMR changes from -3.85% in Sample W2 at 190 K to -7.68% in Sample W1c at 165 K (see Table S1 for results of other samples). The variations, particularly those in samples patterned on the same wafer, i.e., Samples W1a, W1b and W1c, suggest that the NLMR may originate from local properties such as microscopic disorder.

The features including both the monotonic and non-monotonic magnetic field dependences in the $R(B)$ curves shown in Fig.1e are akin to those reported in crystals [9,10,14,16,17] and films [21,22] of topological materials. The $R(B)$ curves presented in Fig.1c for Sample W1b at 170 K and various angles indicate that negative MRs only occur when the magnetic field is aligned within a few degrees to the current flowing direction and become most pronounced at $B//I$, akin to those attributed to chiral anomaly [9,10]. However, the temperature dependence of the NLMR in our quantum wells, as shown in Fig.1d for Sample W1b, differs significantly from those reported in literature, where NLMR diminishes monotonically with increasing temperature [9,10,21,22]. In contrast, the NLMR in our GaAs quantum wells shows a non-monotonic temperature behaviour and becomes most pronounced at ~170 K. This difference could be understood with the existence of two types of microscopic disorders in quantum wells. In addition to the conventional microscopic disorder due to impurity and lattice defects that is relevant at low temperatures, the interface roughness of a quantum well may also induce NLMRs at high temperatures.

Since the samples are in a standard Hall-bar geometry defined through photolithographic patterning and the current contacts are far away from the voltage probes, the NLMRs observed here are unlikely to be artifacts arising from non-uniform current injection [33,39]. In general, artifacts become more pronounced with increasing mobility and thus can be excluded with the non-monotonic temperature behaviour of the observed NLMRs, which differs from the monotonic temperature dependence of the mobility (see Fig.2b and Fig.S5d). The quantum wells are of very high quality, as demonstrated by the Shubnikov de Haas quantum oscillations at $T = 3$ K (see Fig.S6). Their mobility can be larger than $3 \times 10^6$ cm$^2$V$^{-1}$s$^{-1}$ at $T = 3$ K and reach as high as $1.4 \times 10^4$ cm$^2$V$^{-1}$s$^{-1}$ even at room temperature (see Fig.S5d). Therefore, macroscopic defects such as exotic phases and grain boundaries cannot be the origin of NLMRs in the quantum wells.

On the other hand, it is known that microscopic disorder plays a crucial role in the occurrence of quantum Hall plateaus [40], which is observed in all of our quantum wells (see Fig.S6). That is, areas with microscopic disorder exist in the quantum wells alongside with clean areas (see Fig.3a for a schematic). Computer simulations for polycrystalline Ag$_{2\pm x}$Se samples [25,26] have demonstrated that macroscopic disorder such as microscale clusters of excessive Ag induces distorted current paths, resulting in NLMRs. Similarly, the current paths in areas with microscopic disorder are expected to be distorted due to local conductivity fluctuation. Thus, we postulate that the disordered areas where the current paths are distorted in quantum wells have positive (negative) longitudinal magnetoconductances (magnetoresistances). Considering that a quadratic field dependence of the magnetoconductance is the most common relationship revealed by experiments and theories to describe NLMRs, we assume that the magnetotransport behaviour of the disordered areas to be $R_d(B) = R_{d0}/(1 + \alpha B^2)$, where $R_{d0}$ is the resistance of the disordered areas at zero magnetic field. For the magnetoresistance of the clean areas where the current paths are not distorted, we take the conventional Drude form $R_c(B) = R_{c0}(1 + \beta B^2)$ [27]. Figure 3b presents schematics for the expected $R(B)$ relationship for these two scenarios.

The above hypotheses allow us to use a phenomenological three-resistor model to qualitatively account for the various magnetic field behaviours of the observed NLMRs. For a sample with disordered areas mostly surrounded by clean areas (see a schematic in Fig.3a), its $R(B)$ behaviour can be evaluated using a simplified equivalent circuit consisting of three resistors (see Fig.3c), i.e., a resistor $R_c^p$ with positive MR (representing the clean areas) in parallel with two resistors $R_d$ and $R_c^s$ that are in series and have opposite MRs, where $R_d$ and $R_c^s$ are the resistances of the disordered and clean areas, respectively. The superscripts $s$ and $p$ denote the cases of the clean area in series with and in parallel to the disordered area. In this case the $R(B)$ behaviour can be described as

$$R(B) = [1/(R_d + R_c^s) + 1/R_c^p]^{-1} \tag{1}$$

Since each of $R_d$, $R_c^s$, and $R_c^p$ has two variables, Eq.1 has six free parameters. In order to improve the reliability of the analysis, we can take advantage of the measured zero-field resistance by rewriting Eq.1 as

$$R(B) = R_0\{\varepsilon_d/[\gamma_d/(1 + \alpha B^2) + (1 - \gamma_d)(1 + \beta^s B^2)] + (1 - \varepsilon_d)/(1 + \beta^p B^2)\}^{-1} \tag{2}$$

where $R_0 = [1/(R_{d0} + R_{c0}^s) + 1/R_{c0}^p]^{-1}$ is the measured zero-field resistance, $\varepsilon_d = R_{c0}^p/(R_{c0}^p + R_{d0} + R_{c0}^s)$ is the ratio of the conductance of the channel with disordered area to the total value at zero-field and $\gamma_d = R_{d0}/(R_{d0} + R_{c0}^s)$ is the ratio of the zero-field resistance of the disordered area to the corresponding total value of the channel consisting of the disordered and clean areas in series.

We use Eq.2 to account for the experimental $R(B)$ curves as presented in Fig.1e as solid curves. We found that all five fitting parameters are necessary to describe $R(B)$ curves with positive MR near zero-field, like those obtained at $T$ = 3 K and 133 K, resulting in large uncertainty in the analysis. Numerical results in Fig.S7 indicate that $R_c^p$ is the contributor for the positive MR near zero-field (see Fig.S7c) and the experimental $R(B)$ curves at $T \geq 138$ K in Fig.1e follow the NLMR behavior modelled with an equivalent circuit with $R_d$ and $R_c^s$ in series (see Fig.S7b). In this scenario, $\varepsilon_d = 1$, reducing Eq.2 to $R(B) = R_0[\gamma_d/(1 + \alpha B^2) + (1 - \gamma_d)(1 + \beta^s B^2)]$. Using the experimentally determined $R_0$, this simplification decreases the number of variables to three ($\gamma_d$, $\alpha$, and $\beta^s$). This simple serial scenario accounts for the experimental data very well, as demonstrated in Fig.1e for $T$ = 138 K, 150 K and 250 K. Figure 4 shows the temperature dependence of the derived $\gamma_d$, $\alpha$, and $\beta^s$. It is reasonable for $\alpha$ and $\beta^s$ to increase with decreasing temperature, since they should reflect the temperature behaviour of the electron mobility (see Fig.2b). The combined effect of $\gamma_d$, $\alpha$, and $\beta^s$ leads to an enhanced NLMR with decreasing temperature. Between 150 K and 160 K, however, the NLMRs as well as $\gamma_d$, $\alpha$, and $\beta^s$ change their temperature behaviour, resulting in minima in the NLMR versus temperature curves (see Fig.1d) and peaks in those of the three fitting parameters.

We note that the change in the temperature behaviour of NLMRs as well as $\gamma_d$, $\alpha$, and $\beta^s$ coincides with a transition (at $T_p \sim 145$ K) from semiconducting-like to metallic temperature dependence in the zero-field $R_0(T)$ curve and a kink in the $R_{xy}$ curve at $B = 9$ T (see Fig.2a). As shown in Fig.2b, the temperature dependences of both the electron mobility and density also change significantly at 140 K-150 K. Quantitatively, the temperature dependence of the electron density for Sample W1b at $T > 140$ K can be well described with $n = N_0 \exp(-E_A/k_B T)$, with $k_B$ the Boltzmann constant, $N_0 = 3.05 \times 10^{12}$ cm$^{-2}$, and $E_A \approx 34$ meV (see Table S1 for $N_0$s and $E_A$s of other samples). The $E_A$ value falls in the range of the activation energy of 6-200 meV for Si dopants in the Al$_x$Ga$_{1-x}$As layer (x = 0~0.33) [41]. The value of $N_0$ is close to the total Si density of $5.1 \times 10^{12}$ cm$^{-2}$. Thus, the transport at high temperatures is determined by electrons from the Si dopants.

Both experiments [25,26] and simulations [25,26,42] reveal that (quasi-)linear MRs at $B \perp I$ can appear in polycrystalline materials where NLMRs are detected. As presented in Fig.5a for sample W1b at $T = 170$ K and various magnetic field orientations, our quantum wells also exhibit (quasi-)linear MRs at $B \perp I$ and NLMRs at $B//I$. Furthermore, the linearity of the MRs for $B \perp I$ becomes more pronounced as the NLMRs for $B//I$ increase, as demonstrated by the $R(B)$ curves obtained at various temperatures shown in Fig.5b. The observation of (quasi-)linear MR at $B \perp I$ further attests to the role of disorder on the magnetotransport of our quantum wells. As demonstrated in Fig.5c for the $R(B)$ curves obtained at $T = 105$ K, the MR at $B//I$ can be positive and even quasi-linear. In this case, (quasi-)linear MRs may occur for all field orientations.

**Discussion**

Our phenomenological three-resistor model and Eq.2 can be readily applied to account for the NLMRs in films of other materials, in which both micro- and macroscopic disorders are present. The longitudinal magnetoconductance $\sigma_{xx}$ in thin films of topological insulator Bi$_2$Se$_3$ [21] is found to follow $\sigma_{xx} \sim B^2$ in magnetic fields up to 30 T, indicating that the entire sample is disordered. The behaviour of NLMRs in films of Dirac semimetal Cd$_3$As$_2$ is thickness-dependent [22], with various types of $R(B)$ relationship resembling those in Fig.S7. In the scenario of Fig.S7c for a parallel equivalent circuit, Eq.2 can be re-written as $\sigma(B) = \sigma_{d0}(1 + \alpha B^2) + \sigma_{c0}/(1 + \beta^p B^2)$, which is exactly the same equation used in the analysis of the data for the 370 nm thick $Cd_3As_2$ film [22]. On the other hand, the two thinner samples (with thicknesses of 85 nm and 120 nm) exhibit similar type of NLMRs to our quantum wells (of 40 nm thick) at high temperatures, which can be described with the reduced form of Eq.2 for a series equivalent circuit (see Fig.S7b). For films of intermediate thicknesses (170 nm and 340 nm), Eq.2 with all three resistors (see Fig.S7a) is required to account for the observed NLMRs.

Recently reported NLMRs in single crystals were often attributed to chiral anomaly. In quantitative analyses [16,34], possible weak anti-localization conductance $\sigma_{WAL}$ and conventional Fermi surface contribution $\sigma_N$ were also included to account for the deviation from quadratic field dependence of the conductance, leading to

$$\sigma(B) = \sigma_{WAL}(1 + C_W B^2) + \sigma_N \qquad (3)$$

with the positive constant $C_W$ reflecting the contribution from chiral anomaly and $\sigma_{WAL} = aB^{1/2} + \sigma_0^{wal}$ with $a < 0$ and $\sigma_N^{-1} = \rho_0^N + bB^2$. Such an expression with five variables could describe the experimental data at low magnetic fields [16,34]. Since weak anti-localization may also exist in quantum wells at low temperatures, we applied Eq.3 to analyze our data obtained at 3 K and the result is presented in Fig.S2b. With four fitting parameters ($a = 1.98 \times 10^{-3}$ T$^{-2}$, $\sigma_0^{wal} = 2.06 \times 10^{-3}$ Ω, $C_W = 59.84$ T$^{-2}$, $b = 87.21$ T$^{-2}$) and the measured zero-field resistivity $\rho_0$ (= 19.81 Ω) (which gives $\rho_0^N = 1/\rho_0 - \sigma_0^{wal} = 4.84 \times 10^{-2}$ Ω), Eq.3 can indeed describe the results at low magnetic fields ($B < 0.5$ T) well and also reproduces the upturn in the MR at high magnetic fields. However, when we set $C_W = 0$ for our non-Weyl quantum well, Eq.3 reduces to $\sigma(B) = \sigma_{WAL} + \sigma_N$, and yields only positive MR since both $\sigma_{WAL}$ and $\sigma_N$ decrease with increasing magnetic field. Thus, Eq.3 is not applicable to a non-Weyl system. We also used Eq.3 to reveal the possible contribution of weak localization by allowing $a > 0$. In this case, Eq.3 cannot describe the experimental data even at low magnetic fields if we set $C_W = 0$. Thus, weak localization that may induce NLMRs in a 2D system at low temperatures [43-45] cannot be the dominant contributor to the observed NLMRs at low temperatures.

Our results reveal that linear MRs can co-exist with NLMRs. Thus, NLMRs can be used to distinguish disorder based origin from other mechanisms [46-49] for linear MRs. On the other hand, (quasi-)linear MRs for $B \perp I$ could be an indicator of the existence of NLMRs, providing guidance in the search for materials with NLMRs. As demonstrated in disorder-tuned polycrystalline samples [50], our discovery of the role of microscopic disorder on the occurrence of NLMRs and linear MRs will stimulate more work to tune the magnetotransport properties of single crystals through microscopic disorder engineering, e.g., doping [16,20] and irradiations with light charge particles such as protons [51] and electrons [52] for crystals.

**Methods**

**Sample preparation**. The measured samples are 40-nm wide GaAs quantum well grown by molecular beam epitaxy [4]. The quantum well is buried 180 nm deep under the surface, and separated by 150 nm $Al_{0.24}Ga_{0.76}As$ thick barriers on both sides from the $\delta$-doped silicon impurities with densities of $1.18 \times 10^{12}$ cm$^{-2}$ and $3.92 \times 10^{12}$ cm$^{-2}$ for the layers under and above the quantum well, respectively. The samples are fabricated into Hall bars with width of $L_y = 50$ µm and voltage lead distance of $L_x = 100$ µm (see Fig.1a) by photolithography. Contacts to the quantum well are made by annealing InSn at 420 °C for 4 minutes. The electron densities of the measured quantum wells at 3 K range from 7.91 to 8.83 ($10^{10}$cm$^{-2}$) (see Table S1).

**Resistance Measurements.** We conducted DC resistance measurements using a Quantum Design Physical Property Measurement System (PPMS-9). We used constant current mode. The results are found to be current-independent (see Fig.S8). The reported data were taken with $I = 0.5$ $\mu$A for all samples. Angular dependence of the resistance was obtained by placing the sample on a precision, stepper-controlled rotator with an angular resolution of $0.05°$. Figure 1b shows the measurement geometry where the angle $\theta$ between the magnetic field $B$ and the current $I$ can be varied. Fig.S9 shows the experimental definition of $\theta = 0°$, i.e. the orientation of magnetic field at $B//I$. In experiments we measure $R(H)$ curves at various fixed temperatures and angles, as demonstrated in Fig.1c and 1e. The magnetoresistance is defined as $MR = [R(H) - R_0)]/R_0$ where $R(H)$ and $R_0$ are the resistance at a fixed temperature with and without the presence of a magnetic field, respectively.

## Acknowledgements

Magnetotransport measurements were supported by the U.S. Department of Energy, Office of Science, Basic Energy Sciences, Materials Sciences and Engineering. Sample fabrication and characterization were supported by the Department of Energy Basic Energy Sciences (Grant No. DE-FG02-00-ER45841), the National Science Foundation (Grants No. DMR 1709076 and MRSEC DMR 1420541), and the Gordon and Betty Moore Foundation (Grant No. GBMF4420). W. Z. acknowledges support from the DOE Visiting Faculty Program and the U.S. National Science Foundation under Grants No. DMR-1808892. M. Su. was supported by the Fulbright Program. J. X. and Z. X. also acknowledge support by the National Science Foundation under Grant No. DMR-1407175.

**Author contributions**

Z. L. X, Y. L. W. and W. Z. designed the experiments; M. K. M., L. N. P., K. W. W., K. W. B, and M. Sh. grew and fabricated the samples. J. X., M. Su, Y. L. W. and Z. X. conducted the transport measurements, J. X., Y. L. W., D. F. J. and W. Z. contributed to data analysis; Z. L. X, W. Z., Y. L. W and W. -K. K wrote the paper. All of the authors reviewed the manuscript.

**References**

1. Daughton, J. M. GMR applications. *J. Magn. Magn. Mater*. **192**, 334-342 (1999).

2. Mirlin, A.D., Wilke, J., Evers, F., Polyakov, D. G., & Woelfle, P. Strong magnetoresistance induced by long-range disorder. *Phys. Rev. Lett*. **83**, 2801-2804 (1999).

3. Khouri, T., Zeitler, U., Reichl, C., Wegscheider, W., Hussey, N. E., Wiedmann, S. & Maan, J. C. Linear magnetoresistance in a quasifree two-dimensional electron gas in an ultrahigh mobility GaAs quantum well. *Phys. Rev. Lett*. **117**, 256601 (2016).

4. Shi, Q., Martin, P. D., Ebner, Q. A., Zudov, M. A., Pfeiffer, L. N. & West, K. W. Colossal negative magnetoresistance in a two-dimensional electron gas. *Phys. Rev. B* **89**, 201301(R) (2014).

5. Ali, M. N. *et al.* Large, non-saturating magnetoresistance in $WTe_2$. *Nature* **514**, 205-208 (2014).

6. Liang, T. *et al.* Ultrahigh mobility and giant magnetoresistance in Dirac semimetal $Cd_3As_2$. *Nat. Mater*. **14**, 280-284 (2015).

7. Shekhar, C. *et al.* Extremely large magnetoresistance and ultrahigh mobility in the topological Weyl semimetal NbP. *Nat. Phys*. **11**, 645-649 (2015).

8. Tafti, F. F. *et al*. Resistivity plateau and extreme magnetoresistance in LaSb. *Nat. Phys*. **12**, 272-277 (2016).

9. Xiong, J. *et al*. Evidence for the chiral anomaly in the Dirac semimetal Na$_3$Bi, *Science* **350**, 413-416 (2015).

10. Huang, X. C. *et al*. Observation of the chiral-anomaly-induced negative magnetoresistance in 3D Weyl semimetal TaAs. *Phys. Rev. X* **5**, 031023 (2015).

11. Li, Q. *et al*. Chiral magnetic effect in ZrTe$_5$. *Nat. Phys*. **12**, 550-554 (2016).

12. Hirschberger, M. *et al*. The chiral anomaly and thermopower of Weyl fermions in the half-Heusler GdPtBi. *Nat. Mater.* **15**, 1161- 1166 (2016).

13. Zhang, C. L. *et al*. Signatures of the Adler–Bell–Jackiw chiral anomaly in a Weyl fermion semimetal. *Nat. Commun*. **7**, 10735 (2015).

14. Wang, L.-X., Li, C. -Z., Yu, D. -P. & Liao, Z. -M. Aharonov–Bohm oscillations in Dirac semimetal Cd$_3$As$_2$ nanowires. *Nat. Commun*. **7**, 10769 (2016).

15. Wang, H. C. *et al*. Chiral anomaly and ultrahigh mobility in crystalline HfTe$_5$. *Phys. Rev. B* **93**, 165127 (2016).

16. Lv, Y. -Y. *et al*. Experimental observation of anisotropic Adler-Bell-Jackiw anomaly in type-II Weyl semimetal WTe$_{1.98}$ crystals at the quasiclassical regime. *Phys. Rev. Lett*. **118**, 096603 (2017)

17. Gooth, J. *et al*. Experimental signatures of the mixed axial-gravitational anomaly in the Weyl semimetal NbP. *Nature* **547**, 324-327 (2017).

18. Arnold, F. *et al*., Negative magnetoresistance without well-defined chirality in the Weyl semimetal TaP, *Nat. Commun.* **8**, 11615 (2016).

19. Li, P. *et al*. Evidence for topological type-II Weyl semimetal WTe$_2$, *Nat. Commun.* **7**, 2150 (2017).

20. Wang, Z. *et al*. Defects controlled hole doping and multivalley transport in SnSe single crystals. *Nat. Commun.* **9**, 47 (2017).

21. Wiedmann, S. *et al.* Anisotropic and strong negative magnetoresistance in the three-dimensional topological insulator $Bi_2Se_3$. *Phys. Rev. B* **94**, 081302(R) (2016)

22. Schumann, T., Goyal, M., Kealhofer, D. A. & Stemmer, S. Negative magnetoresistance due to conductivity fluctuations in films of the topological semimetal $Cd_3As_2$. *Phys. Rev. B* **95**, 241113(R) (2017).

23. Breunig, O. *et al.* Gigantic negative magnetoresistance in the bulk of a disordered topological insulator. *Nat. Commun.* **8**, 15545 (2017).

24. Han, F. *et al.* Separation of electron and hole dynamics in the semimetal LaSb. *Phys. Rev. B* **96,** 125112 (2017).

25. Hu, J. S., Rosenbaum, T. F. & Betts, J. B. Current jets, disorder, and linear magnetoresistance in the silver chalcogenides. *Phys. Rev. Lett.* **95**, 186603 (2005).

26. Hu, J. S., Parish, M. M. & Rosenbaum, T. F. Nonsaturating magnetoresistance of inhomogeneous conductors: Comparison of experiment and simulation. *Phys. Rev. B* **75**, 214203 (2007).

27. Son, D. T. & Spivak, B. Z. Chiral anomaly and classical negative magnetoresistance of Weyl metals. *Phys. Rev. B* **88**, 104412 (2013).

28. Burkov, A. A. Negative longitudinal magnetoresistance in Dirac and Weyl metals. *Phys. Rev. B* **91**, 245157 (2015).

29. Dai, X., Du, Z. Z. & Lu, H-Z. Negative magnetoresistance without chiral anomaly in topological insulators. *Phys. Rev. Lett.* **119**, 166601 (2017).

30. Andreev, A. V. & Spivak, B. Z. Longitudinal negative magnetoresistance and magnetotransport phenomena in conventional and topological conductors, *Phys. Rev. Lett.* **120**, 026601 (2018).

31. Goswami, P. Pixley, J. H. & Das Sarma, S. Axial anomaly and longitudinal

magnetoresistance of a generic three-dimensional metal. *Phys. Rev. B* **92**, 075205 (2015).

32. Gao, Y., Yang, S. A. & Niu, Q. Intrinsic relative magnetoconductivity of non-magnetic metals. *Phys. Rev. B* **95**, 165135 (2017).

33. Liang, S. H., Lin, J. J., Kushwaha, S., Cava, R. J. & Ong, N. P. Experimental tests of the chiral anomaly magnetoresistance in the Dirac-Weyl semimetals $Na_3Bi$ and GdPtBi. *Phys. Rev. X* **8**, 031002 (2018).

34. Kim, H. -J. *et al.* Dirac versus Weyl Fermions in topological insulators: Adler-Bell-Jackiw anomaly in transport phenomena. *Phys. Rev. Lett.* **111**, 246603 (2013).

35. Rullier-Albenque, F., Colson, D. & Forget, A. Longitudinal magnetoresistance in Co-doped $BaFe_2As_2$ and LiFeAs single crystals: Interplay between spin fluctuations and charge transport in iron pnictides. *Phys. Rev. B* **88**, 045105 (2013).

36. Juyal, A., Agarwal, A. & Soumik Mukhopadhyay, S. Negative longitudinal magnetoresistance in the density wave phase of $Y_2Ir_2O_7$. *Phys. Rev. Lett.* **120**, 096801 (2018).

37. Zhao, Y. F. *et al.* Anisotropic magnetotransport and exotic longitudinal linear magnetoresistance in $WTe_2$ crystals. *Phys. Rev. B* **92**, 041104(R) (2015).

38. Wang, Y. J. *et al.* Gate-tunable negative longitudinal magnetoresistance in the predicted type-II Weyl semimetal $WTe_2$. *Nat. Commun.* **7**, 13142 (2016).

39. dos Reis, R. D. *et al.* On the search for the chiral anomaly in Weyl semimetals: the negative longitudinal magnetoresistance. *New J. Phys.* **18,** 085006 (2016).

40. Yennie, D. R. Integral quantum Hall effect for nonspecialists. *Rev. Mod. Phys.* **59**, 781 (1987).

41. Salmon, L. G. & D'Haenens, I. J. The effect of aluminum composition on silicon donor behavior in $Al_xGa_{1-x}As$. *J. Vac. Sci. & Technol. B* **2**, 197 (1984).

42. Parish, M. M. & Littlewood, P. B. Non-saturating magnetoresistance in heavily disordered semiconductors, *Nature* **426**, 162-165 (2003).

43. Gorbachev, R. V. *et al*. Weak localization in bilayer graphene. *Phys. Rev. Lett.* **98**, 176805 (2007).

44. Minkov, G. M. *et al*. Quantum corrections to conductivity: From weak to strong localization. *Phys. Rev. B* **65**, 235322 (2002).

45. Hassenkam, T. *et al*., Spin splitting and weak localization in (110) GaAs/Al$_x$Ga$_{1-x}$As quantum wells, *Phys. Rev.* B **55**, 9298-9301 (1997).

46. Narayanan, A. *et al*. Linear magnetoresistance caused by mobility fluctuations in n-doped Cd$_3$As$_2$. *Phys. Rev. Lett*. **114**, 117201 (2015).

47. Alekseev, P. S. *et al*. Magnetoresistance in two-component systems. *Phys. Rev. Lett.* **114**, 156601 (2015).

48. Song, J. C. W., Refael, G. & Lee, P. A. Linear magnetoresistance in metals: Guiding center diffusion in a smooth random potential. *Phys. Rev. B* **92**, 180204(R) (2015).

49. Thirupathaiah, S. *et al*. Possible origin of linear magnetoresistance: Observation of Dirac surface states in layered PtBi$_2$. *Phys. Rev. B* **97**, 035133 (2018).

50. Hu, J. S. & Rosenbaum, T. F. Classical and quantum routes to linear magnetoresistance. *Nat. Mater*. **7**, 697-700 (2008).

51. Fang, L. *et al*. Doping- and irradiation-controlled pinning of vortices in BaFe$_2$(As$_{1-x}$P$_x$)$_2$ single crystals. *Phys. Rev. B* **84**, 140504 (2011).

52. Prozorov, R. *et al*. Effect of electron irradiation on superconductivity in single crystals of Ba(Fe$_{1-x}$Ru$_x$)$_2$As$_2$ (x = 0.24). *Phys. Rev. X* **4**, 041032 (2014).

# FIGURE 1

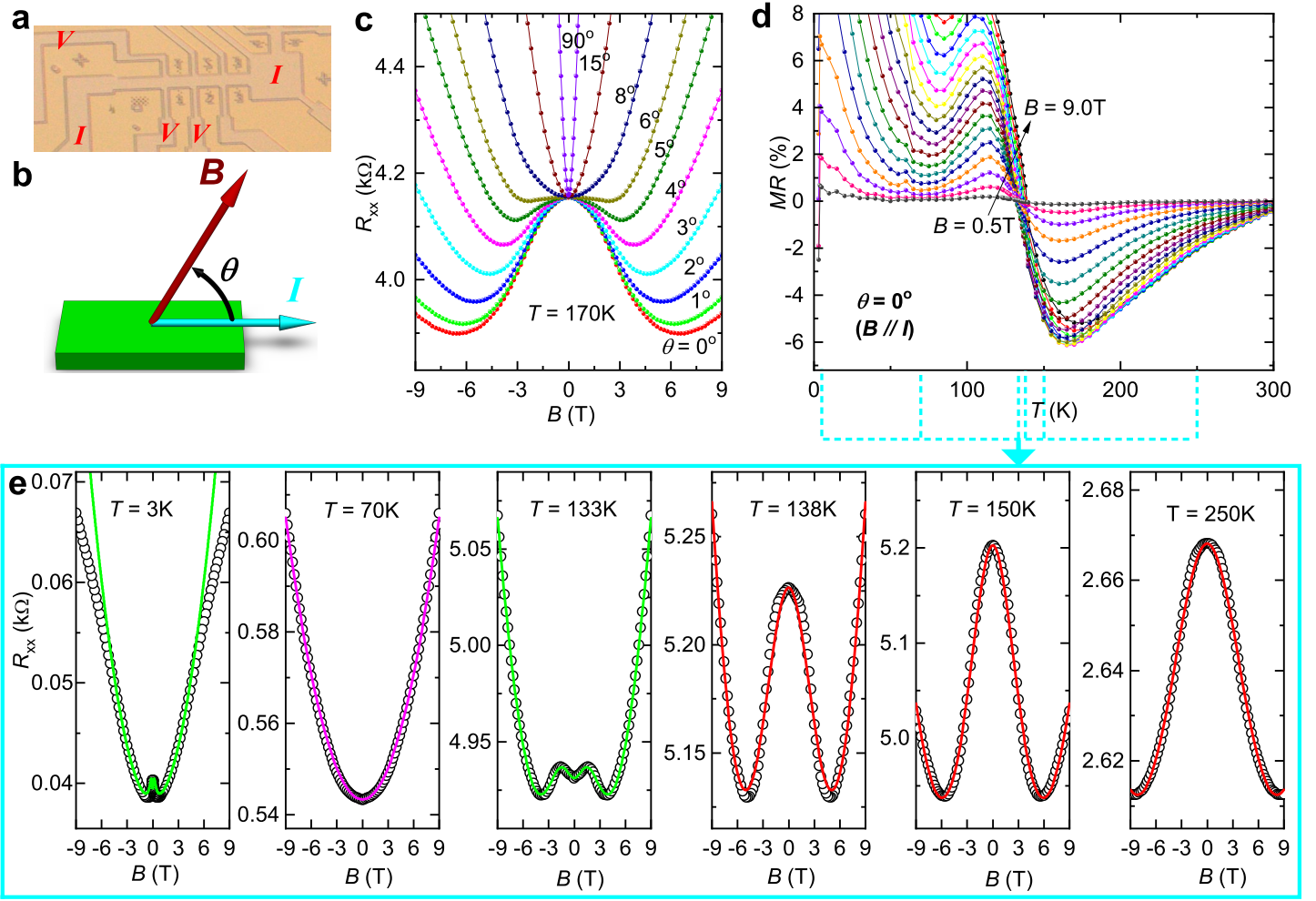

**Figure 1 | Negative longitudinal magnetoresistance in Sample W1b. a,** Micrograph of the sample in Hall bar geometry with width of $L_y = 50$ μm and voltage lead distance of $L_x = 100$ μm. **b,** Schematic showing the definition of the angle $\theta$ for the magnetic field orientation, with $\theta = 0°$ for $B//I$ and $\theta = 90°$ for $B \perp I$. **c,** Magnetic field dependence of the resistance $R(B)$ at various field orientations. Negative magnetoresistance can be clearly seen at $\theta \leq 6°$. **d,** Temperature dependence of the magnetoresistance ($MR$) at magnetic fields from $B = 0.5$ T to $B = 9.0$ T at intervals of 0.5 T at $B//I$, where $MR = [R(B)-R_0]/R_0$ with $R_0$ being the longitudinal resistance $R_{xx}$ at zero field. **e,** Representative $R(B)$ curves showing evolution of the MR feature with temperature. The chosen temperatures are given in the corresponding panels and also marked in (**d**). In (**e**), symbols are experimental data; green lines are fits to the data at $T = 133$ K and 3 K using Eq.2 with values of the five variables of $\varepsilon_d = 0.727$, $\gamma_d = 0.295$, $\alpha = 0.148$ T$^{-2}$, $\beta^s = 6.83×10^{-4}$ T$^{-2}$, and $\beta^p = 0.12$ T$^{-2}$ and $\varepsilon_d = 0.818$, $\gamma_d = 0.21$, $\alpha = 20$ T$^{-2}$, $\beta^s = 0.016$ T$^{-2}$, and $\beta^p = 35$ T$^{-2}$, respectively; red lines (for $T = 250$ K, 150 K and 138 K) are fits with the reduced form of Eq.2 for the serial scenario, with fitting parameters presented in Fig.4, and the magenta line (for $T = 70$ K) describes a quadratic magnetic field dependence $R(B) = R_0 (1+\beta B^2)$ with $\beta = 1.4×10^{-3}$ T$^{-2}$ and the measured $R_0 = 543.4$ Ω.

# FIGURE 2

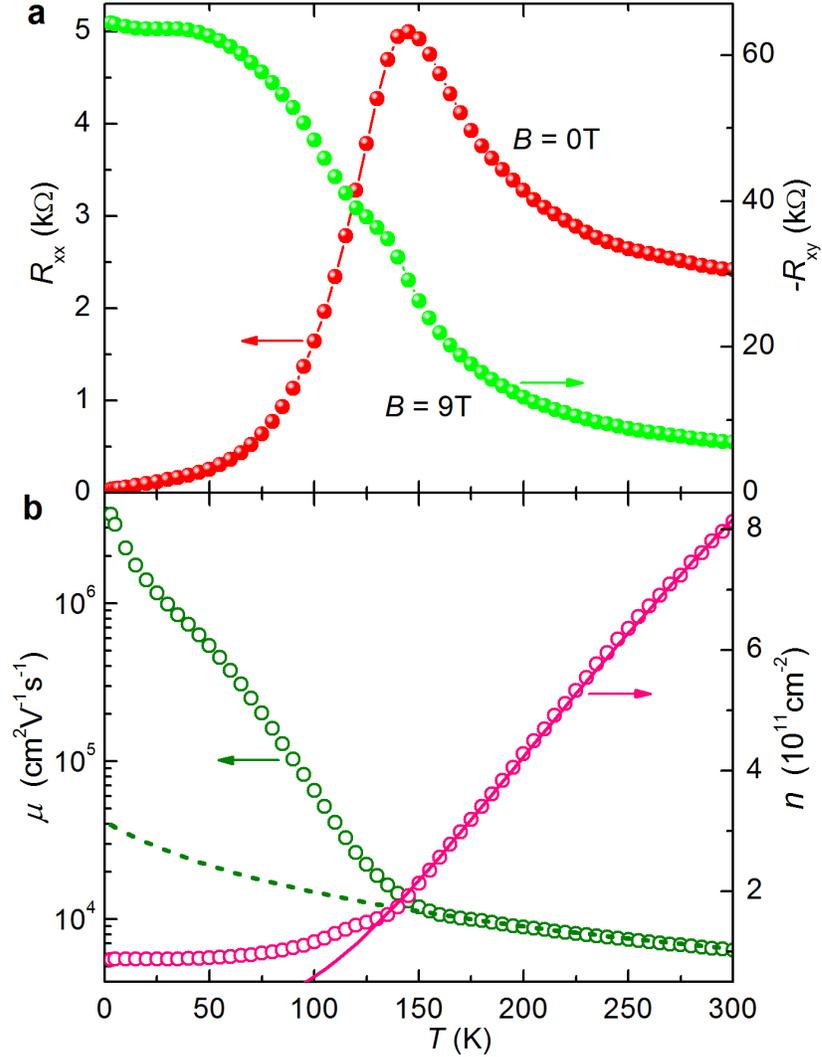

**Figure 2 | Electron density and mobility of Sample W1b. a**, Temperature dependence of the zero-field resistance and the Hall resistance at $B = 9$ T and $\theta = 90°$. **b**, Corresponding electron density and mobility. The electron densities $n$ in (**b**) are calculated from the Hall resistances $R_{xy}$ in (**a**) through the relationship $R_{xy} = B/ne$. The mobilities $\mu$ in (**b**) are derived from the zero-field resistances $\rho_0 = R_{xx}(0T)L_y/L_x$ in (**a**) and the electron densities in (**b**) through the relationship $\rho_0 = 1/ne\mu$. The solid red line in (**b**) is a fit of $n = N_0 \exp(-E_A/k_BT)$, with $k_B$ the Boltzmann constant, $N_0 = 3.05 \times 10^{12}$ cm$^{-2}$, and $E_A = 34.05$ meV. The dotted olive line in (**b**) describes a temperature dependence of the electron mobility $\mu = 2.3 \times 10^6/(55+T)$ (cm$^2$V$^{-1}$s$^{-1}$).

# FIGURE 3

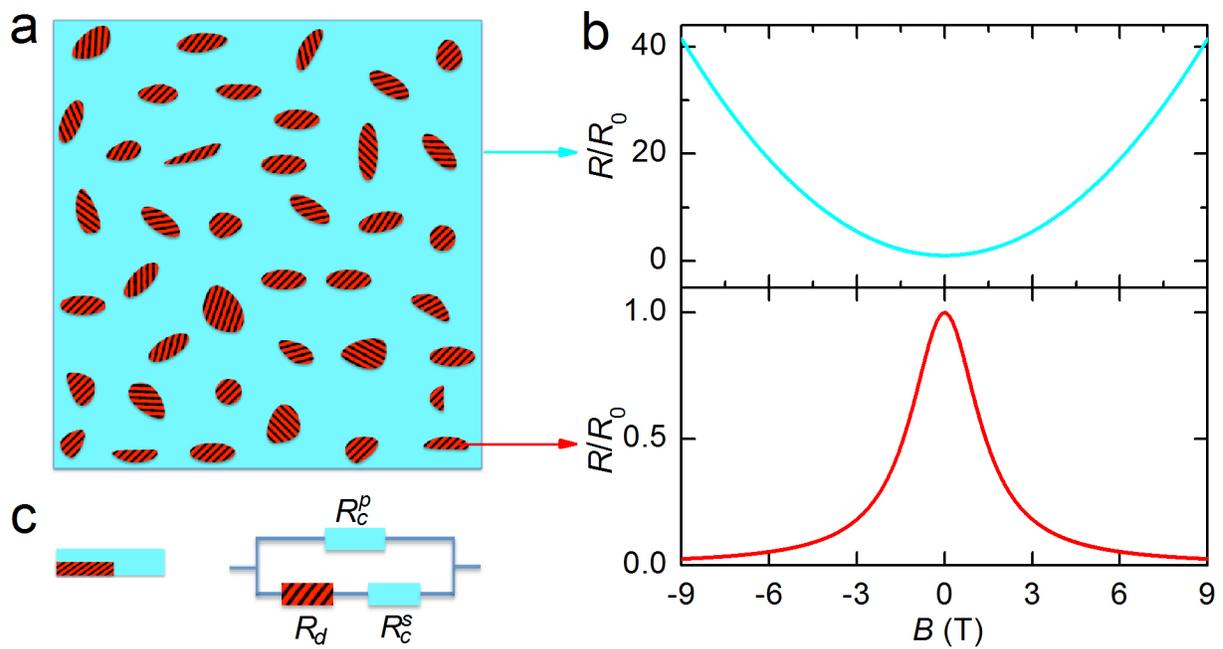

**Figure 3 | Phenomenological three-resistor model. a**, Schematic describing a sample containing areas (red and shaded) where the current paths are distorted and clean areas (cyan) where the current paths are not distorted. **b**, Hypothesized magnetic field dependences of the resistance for the clean (upper panel, $R/R_0 = 1+\alpha B^2$) and disordered regions (lower panel, $R/R_0 = 1/(1+\beta B^2)$ where $\alpha = \beta = 0.5$ T$^{-2}$ are used for the calculations. **c,** Simplified picture for electrotransport in the sample (left panel) and the equivalent circuit (right panel).

**FIGURE 4**

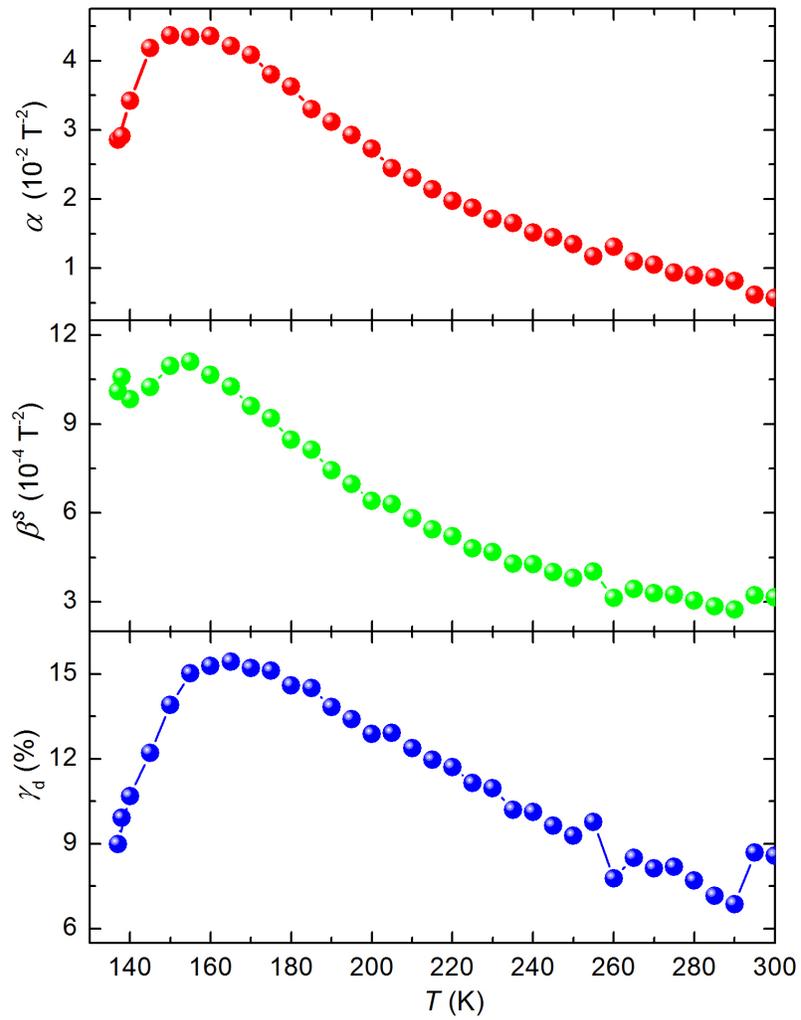

**Figure 4 | Temperature dependence of the derived parameters using Eq.2.** $\alpha$, $\beta^s$ and $\gamma_d$ are defined in the text.

# FIGURE 5

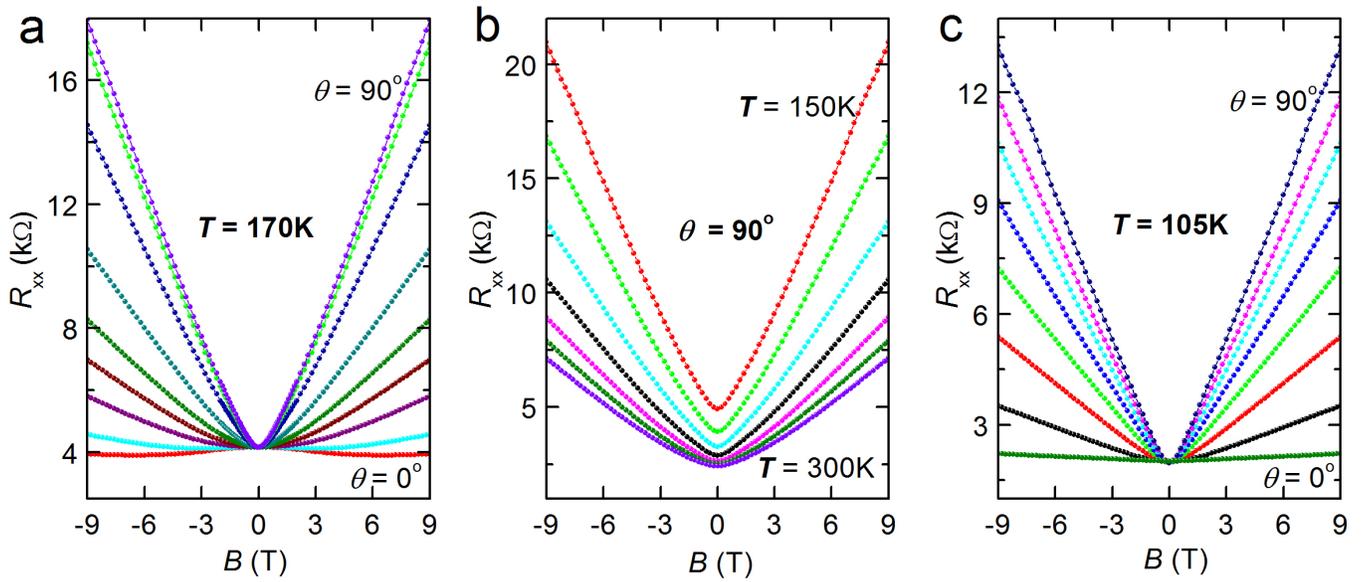

**Figure 5 | Linear magnetoresistances (MR) due to distorted current paths.** **a**, $R(B)$ curves at $T = 170$ K and various magnetic field orientations. Quasi-linear MRs can be seen at high angles. **b**, $R(B)$ curves at $\theta = 90°$ ($B \perp I$) and temperatures from 150 K to 300K at intervals of 25 K. Linearity of the MRs becomes more pronounced with decreasing temperatures at which NLMRs are larger. **c**, $R(B)$ curves at $T = 105$ K and various angles, showing linear MRs in all magnetic field orientations.